\documentclass[preprint,
groupedaddress,
showpacs,
amsmath,amssymb,
aps,
pra,
]{revtex4-1}

\pdfoutput=1

\usepackage[amssymb]{SIunits}
\usepackage{mathrsfs}
\usepackage{graphicx}
\usepackage{dcolumn}
\usepackage{bm}
\usepackage{multirow}
\usepackage{float}
\usepackage{subfigure}

\usepackage{tikz}
\usepackage{cancel}
\usepackage[normalem]{ulem}

\usepackage{xcolor}

\usepackage{booktabs}
\newcolumntype{C}{>{$}c<{$}}
\AtBeginDocument{
\heavyrulewidth=.08em
\lightrulewidth=.05em
\cmidrulewidth=.03em
\belowrulesep=.65ex
\belowbottomsep=0pt
\aboverulesep=.4ex
\abovetopsep=0pt
\cmidrulesep=\doublerulesep
\cmidrulekern=.5em
\defaultaddspace=.5em
}

\begin{document}

\title{Doppler-free saturation of the cascade fluorescence that follows excitation of the $5s \to 6p$ transition in atomic rubidium.}




\author{J.E. Navarro-Navarrete}
\author{A. D{\'\i}az-Calder\'on}
\author{L.M. Hoyos-Campo}
\author{F. Ponciano-Ojeda}
\author{J. Flores-Mijangos}
\author{F. Ram\'irez-Mart\'inez}\email[]{ferama@nucleares.unam.mx}
\author{J. Jim\'enez-Mier}\email[]{jimenez@nucleares.unam.mx}
\affiliation{Instituto de Ciencias Nucleares, UNAM. Circuito Exterior, Ciudad Universitaria, 04510 M\'exico. D.F., M\'exico.}


\date{\today}

\begin{abstract}

We present an experimental scheme that produces Doppler-free spectra of the $5s \to 6p $ second resonance transition in atomic rubidium. The experiment uses the saturation of the cascade fluorescence that occurs when thermal rubidium atoms interact with two counterpropagating $420 $ nm laser beams of comparable intensity. Part of the excited atomic population goes through the $5p_{3/2}$ level which then decays by emission of \unit{780}{\nano\meter} photons. Narrow dips appear in this otherwise broad \unit{780}{\nano\meter} fluorescence, which allow resolution of the $6p_{3/2}$ hyperfine structure. A rate equation model is used to interpret the spectra. It is also shown that these narrow peaks can be used to lock the frequency of the $420 $ nm laser. Using a second beam modulated in frequency produces three sets of spectra with known frequency spacings that can be used to perfom an all-optical measurement of the hyperfine splittings of the $6p_{3/2} $ manifold in rubidium.

\end{abstract}

\maketitle


\section{Introduction}
The second resonant transition in alkali atoms has proven to be very useful in basic and applied research. 
For instance, the $5s \to 6p $ transition in atomic rubidium has provided a confirmation for numerical calculations in magneto-optical and non-lineal rotational models beyond the D1 and D2 lines \cite{Akulshin2014,pustelny_nonlinear_2015}.
It is also an effective, weak EIT probe of a three-level atom in the V configuration \cite{vdovic2007}.
The $5s_{1/2} \to 6p_{3/2}$ transition is commonly used as a first step for the generation of Rydberg atoms \cite{Viteau2011,Valado2013,Bernien2017,Engel2018}. 
The frequencies of the lines in the $5s_{1/2} \to 6p_{1/2}$ absorption spectrum have been measured with uncertainties better than $\unit{70}{\kilo\hertz}$~\cite{Shiner2007}. This precise measurement has led to the $5s \to 6p$ lines to be used as frequency references in the blue range of the visible spectrum \cite{Zhang2017, Glaser2019}. These transitions also form part of a four-level optical clock \cite{Zhang2013}.
The region of the spectrum in which these transitions occur makes them useful in applications in underwater and free space communication systems, and in remote laser monitoring systems \cite{Ling2014}.
One remarkable feature of these transitions is that they are weak, compared with the stronger D1 and D2 lines, which is an advantage in the perturbative study of cold atoms \cite{pustelny_nonlinear_2015}.


The direct excitation of the $6p_{3/2}$ level by means of $\unit{420}{\nano\metre}$ radiation gives place to fluorescence cascade decays which also provide means for very efficient detection \cite{Akulshin2014,pustelny_nonlinear_2015}. 
In addition to the direct decay back to the ground state, the cascade decay may occur via intermediate states ($6s$, $4d$) which then decay to the $5p$ level. In this work we studied the fluorescence signal from the emission at $\unit{780}{\nano\metre}$ that takes place during $5p_{3/2}$ level decay from the $5p_{3/2}$ level to the ground state. Using two counter-propagating beams in the excitation scheme gives place to the generation of Lamb dips in the fluorescence spectra \cite{Sorem1972,She1978,She1995} and measurements of the lineshapes of the D1, D2 transitions in Li \cite{Rowan2013}. 

 Moreover, the use of counter-propagating excitation beams leads to Doppler free spectra, resulting in a simple technique that can be used to resolve the hyperfine structure of the $6p_{3/2}$ manifold for thermal atoms of both rubidium isotopes. 
A simple calculation based on the rate-equation approximation \cite{She1978} which includes the $5s $ ground and $6p_{3/2} $ excited states, and that considers all possible cascade decay paths without polarization is also presented. Furthermore, we show that by using an Acusto-Optic Modulator (AOM) and combining the original beam and modulated beam in a counter-propagating configuration (4 beams interacting with the atomic vapor), we were able to measure the hyperfine splitting of the $6p_{3/2} $ manifold.
Finally, we demonstrate that the Doppler-free fluorescence peaks can be used to lock the excitation laser frequency to any of the observed atomic transitions or cross-over resonances.

Section \ref{Experimental} describes the experimental configuration for thermal atom excitation and fluorescence emission detection. Section \ref{Calculation} presents calculation details for the Doppler free saturation effect. 

Einstein rate equations regarding atomic population of all laser field coupled hyperfine states and also all intermediate decay states will be considered. Furthermore, by obtaining an expression of the $5p_{3/2}$ depopulation and considering the atomic velocity distribution, a profile simulating the resonant fluorescence saturated experimental spectra is obtained. Section \ref{Results} shows the experimental fluorescence spectra for one and two beams inside the Rb cell. A comparison between the experimental and calculated spectra is presented. A description of the absolute frequency scale calibration based on an acousto-optical modulator for hyperfine splitting measurements is described. A comparison between our all optical results for the hyperfine splittings with the previously reported values is made. It is also shown that these hyperfine resolved spectra can be use to lock the frequency of the excitation laser. Finally conclusions are made in section \ref{Conclusions}.

\section{Experimental setup.}\label{Experimental}

\begin{figure}
   \centering
       \includegraphics[width=0.7 \linewidth]{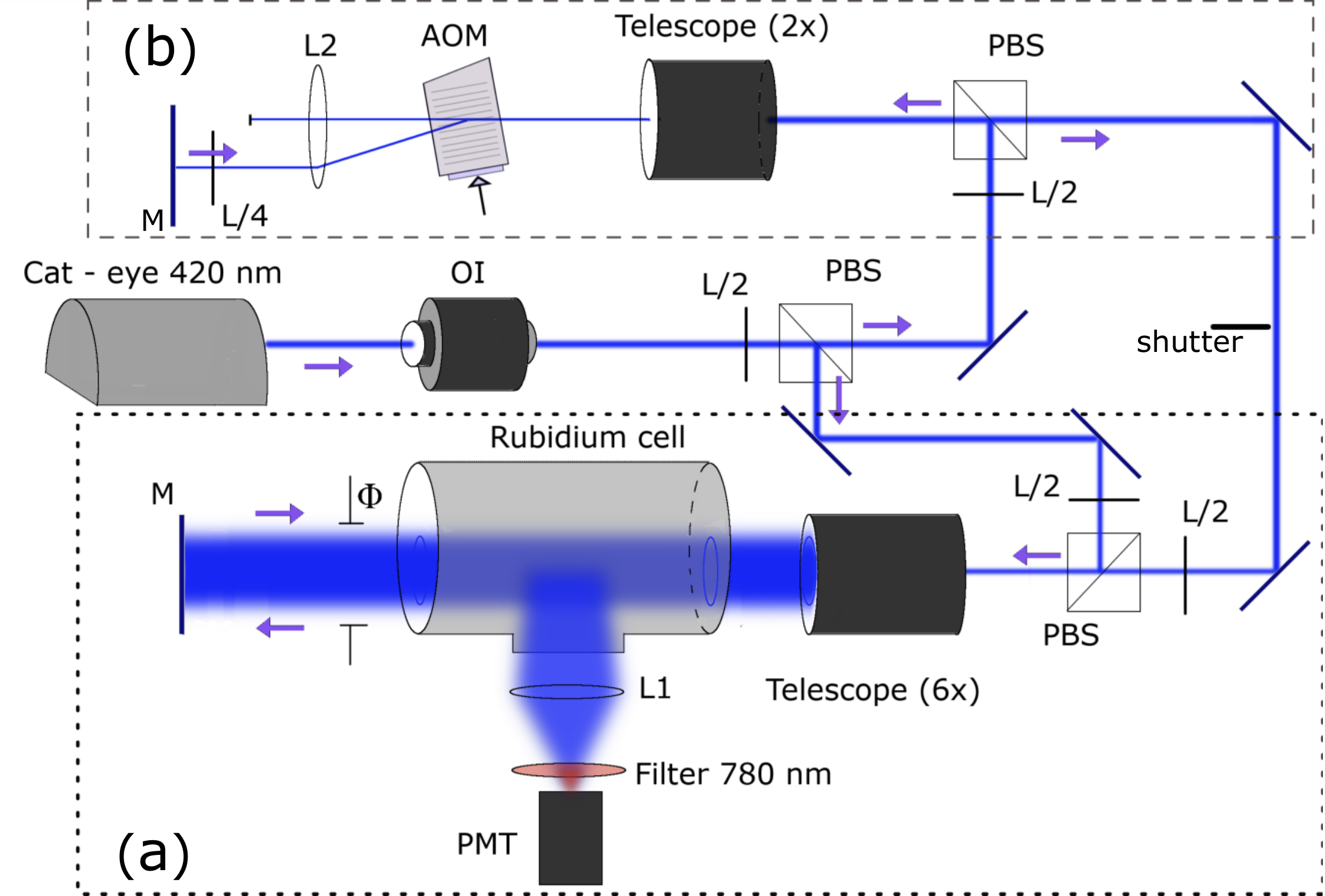}
       \caption{Experimental setup for fluorescence spectroscopy. Cat-eye ECDL: laser of $\unit{420}{\nano\metre}$, OI: Optical Isolator, AOM: Acousto-optic modulator, PBS: polarizing beam splitters, L/2 : half-wave plates, L/4: quarter-wave plate, L1,L2: Plano-convex lenses, PMT: photomultiplier tube; M: flat mirror. The lower part of the experimental setup (a), surrounded by dotted lines, corresponds to the fluorescence generation and detection. The upper part (b) of the setup, enclosed in dash lines, involves the AOM double pass configuration setup. The shutter is used to separate the experiments described in the text.}
       \label{ExpSetup}
 \end{figure}

The experimental setup is shown in figure \ref{ExpSetup}. The same experimental arrangement was used to perform two separate experiments. In the first one (a), a narrow bandwidth diode laser (Moglabs ECDL with emission at $\unit{420}{\nano\metre}$) is used to excite rubidium atoms in a room temperature glass cell with a magnetic µ-metal layer. The laser beam is expanded ($6 \times$) in a telescope and then sent through the glass cell.  A flat mirror is used to retroreflect the $\unit{420}{\nano\metre}$ beam back into the rubidium cell. The polarization of both beams is vertical. The $\unit{780}{\nano\metre}$ cascade fluorescence is detected in the direction perpendicular to the propagation of the excitation beam by means of a photomultiplier tube (PMT, Hamamatsu H5784). A narrow bandpass filter centered at $\unit{780}{\nano\metre}$ with a bandwidth of $\unit{3}{\nano\metre}$ and a 4f lens system are used to collect and focus this fluorescence into the cathode of the PMT. Cascade fluorescence spectra are recorded with and without the retroreflected blue beam.  

In the second experiment, a portion of the $\unit{420}{\nano\metre}$ laser beam is sent through an AOM (AA optoelectronic) in a double pass configuration \cite{Donley2005} ((b) in fig. \ref{ExpSetup}). The main motivation for adding this feature to the spectroscopy system is to provide an absolute frequency calibration and precise control of the excitation laser detuning with respect to the atomic resonances. 

The two beams previously mentioned will be hereafter referred to as the principal and modulated beams, respectively. Both beams then interact with the rubidium atoms having crossed polarizations. The estimated FWHM diameters of the principal and modulated beams are $\unit{11.8}{\milli\meter}$ and $\unit{13.0} {\milli\meter}$ respectively. The intensity employed is $\unit{23.8}{\micro\watt/mm^2}$ and $\unit{27.1}{\micro\watt/mm^2}$ for the principal and modulated beams respectively. The position of the beams inside the cell was selected to put them closer to the PMT in order to minimize reabsorption of the $\unit{780}{\nano\metre}$ fluorescence within the Rb cell. The enlarged beam area increases the interactions of the atoms with radiation and therefore, the hyperfine signal gets amplified.

\section{Calculation} \label{Calculation}

\begin{figure}
   \centering
       \includegraphics[width=0.5\linewidth]{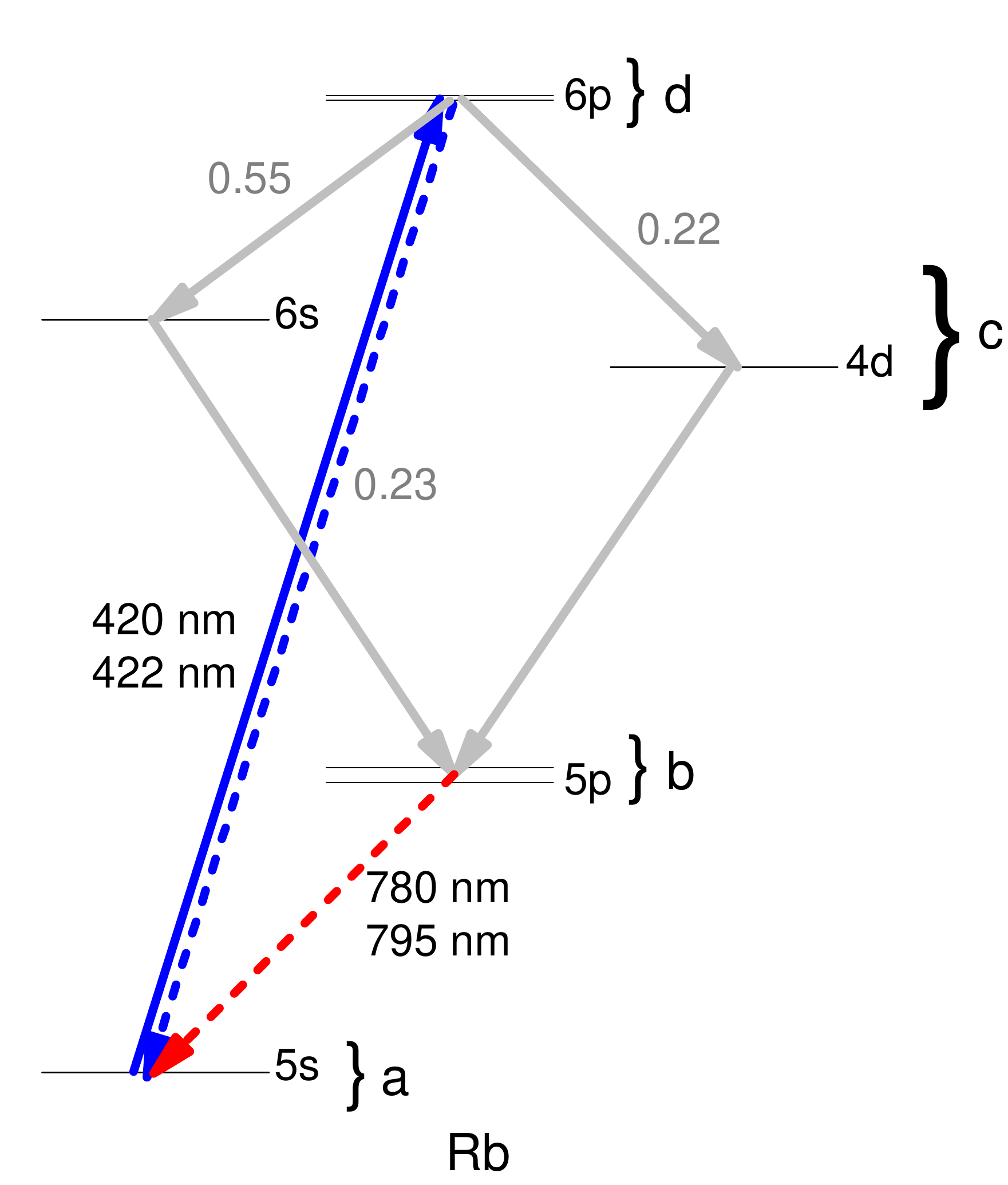}
       \caption{\label{fig:RbEnLvFluo} Rubidium energy level diagram showing in gray the decay for the different energy levels with the corresponding branching ratios \cite{Moon2012}. A $\unit{420}{\nano\metre}$ photon excites the $5s_{1/2} \to 6p_{3/2} $ transition. The excited atoms will then decay back to the ground state either by emission of a blue photon or by any of the two cascades shown in gray. The $5p_j \to 5s_{1/2}$ ($j=1/2,3/2$) transition gives place to the emission of infrared photons at $\unit{780}{\nano\metre}$ or $\unit{795}{\nano\metre}$.}
\end{figure}
%

The principle of the experiments discussed here is better understood with the help of Fig. \ref{fig:RbEnLvFluo}, which shows an energy level diagram of the rubidium atom. A $\unit{420}{\nano\metre}$ photon excites a $5s$ electron into the $6p_{3/2}$ energy level. The excited atom can decay back to the ground state by emission of a $\unit{420}{\nano\metre}$ photon. It can also return to the ground state through the cascades of radiative decays indicated in gray. These cascade decays transit necessarily through the intermediate $5p_j$ levels ($j=1/2,3/2$) and finally decay back down to the ground state with the emission of infrared fluorescence of either $\unit{780}{\nano\metre}$ or $\unit{795}{\nano\metre}$. In this work, we only detected the infrared fluorescence ($5p_{3/2} \rightarrow 5s_{1/2} $) which was selected with a bandpass filter.
Here we adapt the rate equation model used to describe saturation effects in resonant fluorescence of atomic sodium that was reported in Ref. \onlinecite{She1995}.

The calculation considers the entire hyperfine structure of all states in Fig. \ref{fig:RbEnLvFluo} and assumes non-polarized light. 
Calling a, b and d the $5s$, $5p $ and $6p $ fine and hyperfine manifolds, respectively, and denoting by c the $6s $ and $4d $ intermediate manifolds, we construct the equations that describe the population dynamics as follow:  
The $5s_{1/2} $ (a) and $6p_{3/2} $ (d) manifolds are the only ones that are coupled by the laser fields. 
The equations for the excitation dynamics between these states are

\begin{equation}
\frac{dN_d}{dt} = 
\sum_a \left( W_{ad} N_a - W_{da} N_d \right) - \sum_\ell A_{d \ell} N_d - \gamma N_d 
\label{eq:dpop}
\end{equation}
\noindent for the $6p_{3/2} $ hyperfine multiplet, and
\begin{equation}
\frac{dN_a}{dt} = 
\sum_d \left( W_{da} N_d - W_{ad} N_a \right) + \sum_\ell A_{\ell a} N_{\ell} + \gamma\left( \frac{g_a}{g} - N_a\right)
\label{eq:apop}
\end{equation}

\noindent for the $5s_{1/2} $ ground multiplet.
In these equations $W_{ad} $ and $W_{da} $ are, respectively, the absorption and stimulated emission rates. The $A_{ij} $ are the spontaneous emission rates between states $i $ (upper) and $j $ (lower).
The terms with the factor $\gamma $ are included to take into account the entrance and exit of atoms from the overlap region of the laser beams. 
This means that state $i $ loses population at a rate given by $-\gamma N_i $ while the two $5s $ hyperfine states gain population at the rate $\gamma g_a/g $, where $g_a = 2F_a+1 $ is the degeneracy of the hyperfine state and $g $ is the total degeneracy of the $5s $ multiplet.

The populations of all intermediate states (b and c) obey cascade equations of the form
\begin{equation}
\frac{dN_i}{dt} = \sum_k A_{ki} N_k - \sum_{\ell} A_{i\ell} N_i-\gamma N_i
\label{eq:cascade}
\end{equation}
\noindent where the $k $ sum is over all upper states that can decay into state $i $ and $\ell $ represents all states into which state $i $ can decay.

Combining eqs. \ref{eq:dpop}, \ref{eq:apop} and \ref{eq:cascade} a set of 22 (24) rate equations is obtained for all populations of $^{87} $Rb ($^{85} $Rb) up to the $6p_{3/2} $ state. 
These equations can then be solved in the steady state by making $dN_i/dt = 0 $.
The structure of the resulting algebraic equations allows the following stepwise solution.
Equation \ref{eq:dpop} is first solved for the $6p_{3/2} $ populations $N_d $ in terms of the  $5s $ ground state populations $N_a $.
Then the populations of the $6s $ and $4d $ intermediate states are obtained by solving the corresponding equations, of the form of eq. \ref{eq:cascade}, in terms of $N_d $'s. 
The next step of the cascade yields a set of equations that give $N_b $ in terms of $N_c $.
As a result of this procedure the $5p $ populations can be expressed in terms of $N_a $.
Finally, these $N_b $ are substituted into eqs. \ref{eq:apop} which become algebraic equations that must be satisfied by populations $N_a $. 
The inclusion of the entrance/exit terms guarantees the normalization condition
\begin{equation}
\sum_a N_a + \sum_b N_b + \sum_c N_c + \sum_d N_d = 1
\end{equation}

Once one gets all populations in terms of $W_{ad} $, $W_{da}$ and $A_{ij} $, the $780 $ nm fluorescence will be proportional to
\begin{equation}
\phi = \sum_{a,b} N_b A_{ba}
\end{equation}
\noindent where the $b $ sum is only taken over the hyperfine states of the $5p_{3/2} $ manifold.

The values of the spontaneous decay rates $A_{ij} $ are calculated in terms of the transition line strengths $S_{ij} $ according to \cite{She1995}
\begin{equation}\label{EQ:As}
A_{ij} = \frac{\omega_{ij}^3}{3 \pi \epsilon_0 c^3 \hbar (2 F_i +1)} S_{ij}^2
\end{equation}       
\noindent where $F_i $ is the total angular momentum of the upper state, and the line strength of a transition between any two hyperfine states is given by \cite{She1995}:
\begin{equation}\label{EQ:Sij}
{S_{ij}}^2 = (2F_i+1)(2 F_j+1)
\left\{\medspace 
\begin{array}{ccc}
 F_i & 1 & F_j \\
 J_j & I & J_i \\
\end{array}
\medspace \right\}^2 
(S^0_{J_i J_j})^2
\end{equation}
\noindent The values of the $S^0_{J_i J_j} $ line strengths for all states in our rubidium cascade were calculated by Safronova and co-workers \cite{Safronova1999,Safronova2004}.  
For completeness they are also given in Table \ref{tab:Safronova}.

\begin{table}
\caption{Values of the quantities used in the calculation.\label{tab:Safronova}}
\begin{tabular}{@{} c c c c c @{}}
  			\toprule
Upper state & Lower state & $\lambda $ (nm)\footnote{Values calculated with the data of NIST}  & $S (e a_0)$ \footnote{Values taken from references \onlinecite{Safronova1999} and \onlinecite{Safronova2004}} & $A $\footnote{Values calculated with equations \ref{EQ:As} and \ref{EQ:Sij}.} ($MHz$) \\  
\midrule 
$5p_{1/2} $ & $5s_{1/2} $ & 794.979 & 4.221 & 35.9253 \\
$5p_{3/2} $ & $5s_{1/2} $ & 780.241 & 5.956 & 37.8294 \\
$4d_{5/2} $ & $5p_{3/2} $ & 1529.37 & 10.634 & 10.6752 \\
$4d_{3/2} $ & $5p_{1/2} $ & 1475.64 & 7.847 & 9.70663 \\
$4d_{3/2} $ & $5p_{3/2} $ & 1529.26 & 3.54 & 1.77488 \\
$6s_{1/2} $ & $5p_{1/2} $ & 1323.88 & 4.119 & 7.40756 \\
$6s_{1/2} $ & $5p_{3/2} $ & 1366.87 & 6.013 & 14.3428 \\
$6p_{3/2} $ & $5s_{1/2} $ & 420.299 & 0.541 & 1.99676 \\
$6p_{3/2} $ & $4d_{5/2} $ & 2253.58 & 6.184 & 1.6925 \\
$6p_{3/2} $ & $4d_{3/2} $ & 2253.8 & 2.055 & 0.186845 \\
$6p_{3/2} $ & $6s_{1/2} $ & 2732.18 & 13.592 & 4.58823 \\
\bottomrule
\end{tabular} 
\end{table}

The absorption ($W_{ad} $) and stimulated emission ($W_{da} $) rates are expressed in terms of the laser detuning $\delta = \nu - \nu_{ij} $, the laser intensity $I_l $  and the line strength $S_{da} $ according to \cite{She1995}:
\begin{equation}
g_a W_{ad} = g_d W_{da} = \frac{\pi}{3 \epsilon_0 c \hbar^2} S_{ad}^2 \ G(\delta,\Gamma) I_l
\end{equation}  
\noindent Here $G(\delta,\Gamma) $ is the line profile
\begin{equation}
G(\delta,\Gamma) = \frac{1}{2 \pi^2} \frac{\Gamma_{6P}/4 \pi}{\delta^2+(\frac{\Gamma_{6P}}{4 \pi})^2}
\end{equation}
\noindent The line width used in this line profile is the total decay rate of the $6p_{3/2} $ state, which is $\Gamma_{6p} = \sum A_{6p_{3/2}} = 2 \pi \times 1.347 $ MHz.

To include the velocity distribution of the atoms, one writes the detuning in the rest frame of the atom in the presence of two counterpropagating laser beams of equal intensity, which gives a line profile
\begin{equation}
G(\nu,v,\Gamma) = \frac{\Gamma_{6P}}{8 \pi^3} \left[\frac{1}{ \left(\nu-\nu_{da} - \frac{v}{\lambda} \right)^2 +(\frac{\Gamma_{6P}}{4\pi})^2} +\frac{1}{ \left(\nu-\nu_{da} + \frac{v}{\lambda} \right)^2 +(\frac{\Gamma_{6P}}{4\pi})^2} \right]
\label{eq:sfDoppler}
\end{equation}
\noindent Hence, obtaining a $780 $ nm fluorescence intensity that results from excitation with a laser of frequency $\nu $ interacting with an atom with velocity $v $ along the laser propagation direction, $\phi(\nu,v) $. 
The fluorescence observed in the laboratory is then proportional to
\begin{equation}
\Phi(\nu) = \sqrt{\frac{m}{2 \pi k_B T}} \int \phi(\nu,v) \exp\left[-\frac{m v^2}{2 k_B T}\right] dv
\label{eq:finalfluo}
\end{equation}

The evaluation of this fluorescence function (eq. \ref{eq:finalfluo}) was performed using the program Mathematica. 
The population of the $5p_{3/2} $ states were obtained using numerical values of all decay rates. Using the functional dependence of eq. \ref{eq:sfDoppler}, a numerical integration over the Doppler profile was carried out. 
An entrance/exit rate of $0.1 $ MHz was used. 
This corresponds to an average transit time of the atoms across the laser beams of $5 \ \mu $s.      

%
%
%
%

\section{Results and Discussion} \label{Results}
%
\begin{figure}
   \centering
       \includegraphics[width=0.8\linewidth]{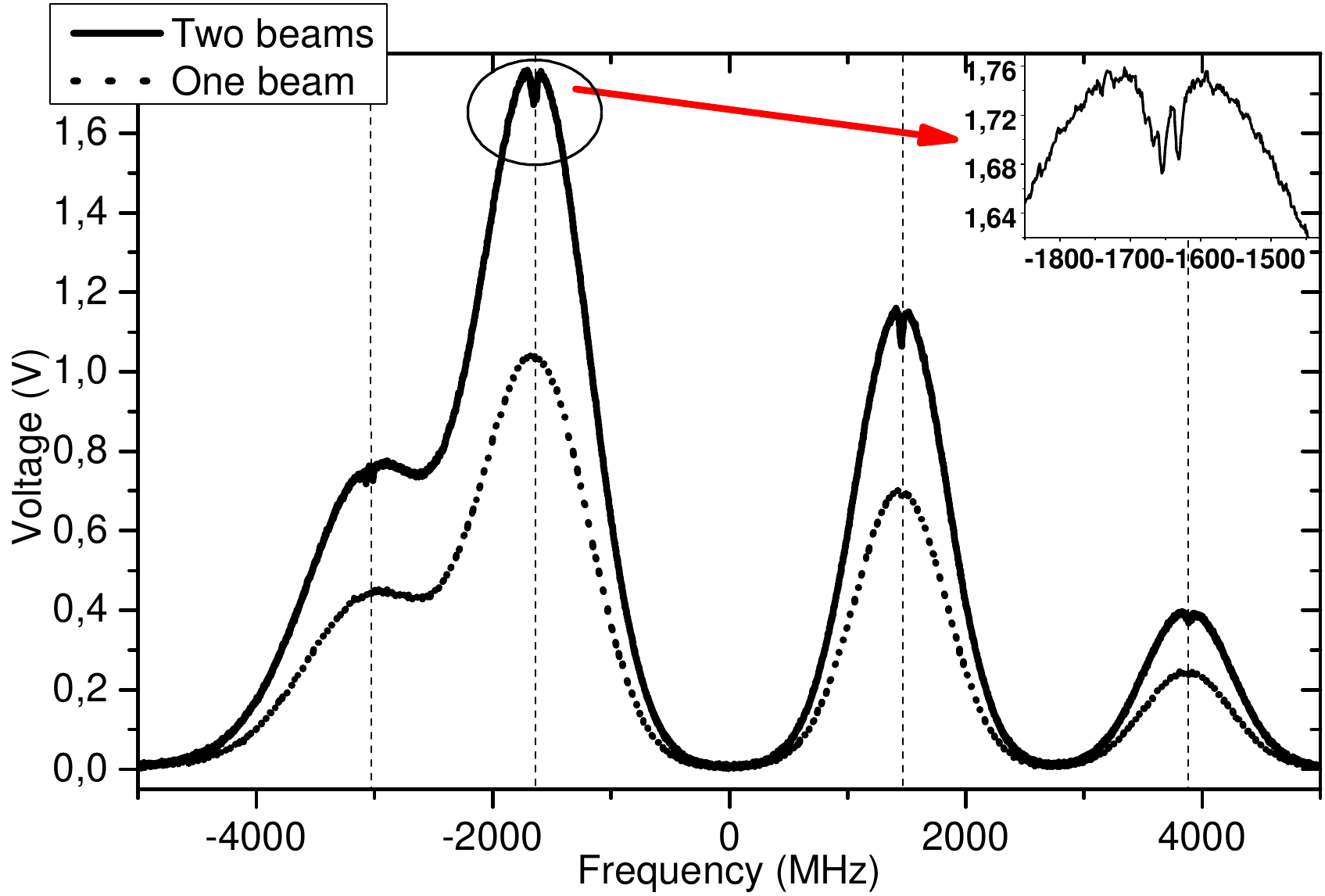}
       \caption{Spectral signal with the incidence of one or two $\unit{420}{\nano\metre}$ beams. The spectra shows four wells corresponding, from left to right, to $5s_{1/2} F=2 \rightarrow 6p_{3/2} F=1,2,3$, $5s_{1/2} F=3 \rightarrow 6p_{3/2} F=2,3,4$ of the high hyperfine ground states of $^{87}Rb$ and $^{85}Rb$ respectively, and $5s_{1/2} F=2 \rightarrow 6p_{3/2} F=1,2,3$, $5s_{1/2} F=1 \rightarrow 6p_{3/2} F=0,1,2$ of the low hyperfine ground states of $^{85}Rb$ and $^{87}Rb$ respectively.}
       \label{OneAndTwoBeams1}
 \end{figure}

A comparison between the $\unit{780}{\nano\metre}$ cascade fluorescence emission recorded with and without a retro-reflected $\unit{420}{\nano\metre}$ beam is shown in Figure \ref{OneAndTwoBeams1}. 
As described in section \ref{Calculation}, this fluorescence emission spectra is generated as a result of atomic excitation population to the $6p_{3/2}$ level and subsequent decay via the $5p_{3/2}$ level. 
Both measurements show four Doppler broadened peaks, whose heights in the spectrum with two beams is roughly twice as that of the peaks in the spectrum recorded with only one beam. This happens because, at a given laser frequency, the two counter-propagating beams interact with two separate groups of atomic velocities within the Doppler width and each independently contributes to the recorded fluorescence signal. 
Furthermore, the two-beam spectrum also shows saturation effects in the center of each Doppler feature, where the two light components simultaneously interact with the same atomic velocity groups.
The inset in Figure \ref{OneAndTwoBeams1} provides a closer look at one of these regions, specifically the one corresponding to the $F=3 \to F'= 2, 3, 4$ set of hyperfine transitions in $^{85}$Rb. There, dips caused by the suppression of the cascade emission reveal the hyperfine structure of the $6p_{3/2}$ level with sub-Doppler resolution, which is clearly not the case when only one beam is crossing through the cell.

The presence of the counter-propagating beam, when this and the principal beam are simultaneously resonant with a same group of atomic velocities, induce the stimulated emission from the $6p_{3/2}$ level directly to the ground level.
This inhibits spontaneous decays through alternate channels, partially depopulating the $5p_j$ levels and therefore reducing fluorescence emission at $\unit{780}{\nano\metre}$. Hence, dips of diminished fluorescence of $\unit{780}{\nano\metre}$ appear in the fluorescence spectra. Therefore, the counter-propagating beam configuration gives rise to a method for registering the hyperfine structure of the $6p_{3/2}$ level with sub-Doppler resolution.

\begin{figure}
 	\centering
 	{\includegraphics[width=\linewidth]{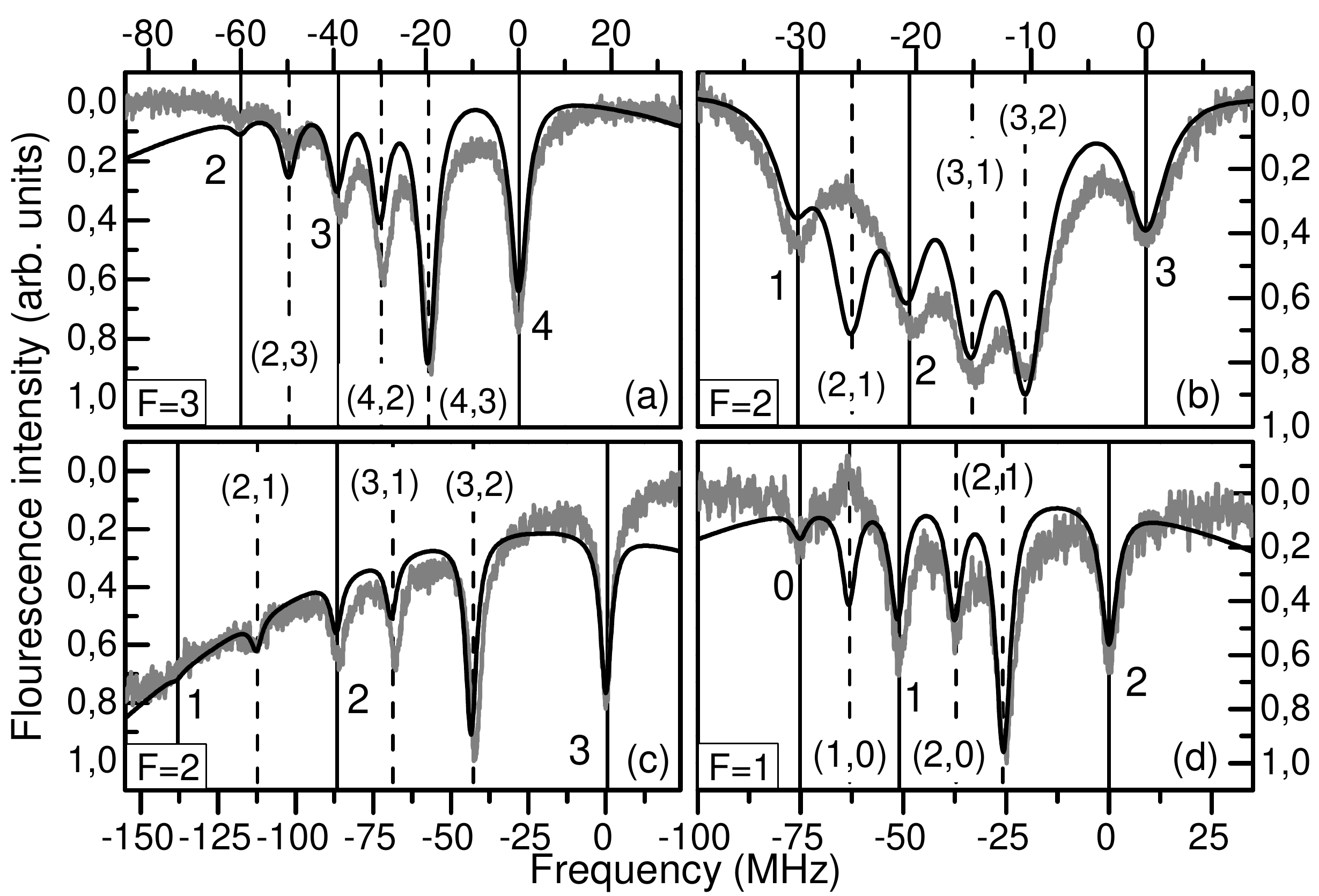}}
 	\caption{The $6p_{3/2}$ level hyperfine splitting for (a,b)  $^{85}$Rb and (c,d) $^{87}$Rb. The solid lines are showing the results from the calculation and the grey line shows the experimental results. The dashed vertical lines are showing the crossovers positions and the vertical solid lines are showing the atomic transition.}
 	\label{OnebeamTwobeams}
 \end{figure}

Figure \ref{OnebeamTwobeams} shows a typical set of spectra of the fluorescence signal recorded as the $\unit{420}{\nano\metre}$ lasers frequency is scanned through the $6p_{3/2}$ manifold. The three expected atomic excitation lines originating from the hyperfine levels of the ground state of each isotope are clearly resolved, ($F=3 \to F'= 2, 3, 4$) and ($F=2 \to F'= 1, 2, 3$) for $^{85}$Rb, presented in graphs (a) and (b) respectively; ($F=2 \to F'= 1, 2, 3$) and ($F=1 \to F'= 0, 1, 2$) for $^{87}$Rb, respectively shown in graphs (c) and (d). The relative positions of these atomic resonances are indicated with vertical solid lines in the graphs. Three additional lines resulting from crossovers, are also well resolved in all four sections of the spectra. These in turn are indicated in the graphs by means of dashed vertical lines.
The excellent agreement between the experimental data and the results from the calculation is shown. The expected atomic and crossovers transitions are in agreement with the obtained experimental results except for the crossover (2, 1) for $^{85}$Rb in (b) and the crossover (1, 0) for $^{87}$Rb in (d). The authors attribute these differences to light polarization effects that require further research. The plot (c) shows a different slope with respect to the other three plots; this is due to the closeness of the $^{85}$Rb F=3 well that influence the trend of the hyperfine structure of the $^{87}$Rb F=2. 


The spectra shown in Fig. \ref{OnebeamTwobeams} can be used for locking the frequency of the \unit{420}{\nano\metre} laser to any of the observed atomic or crossover transitions. As an example, when locked at the half-maximum point of the $F=3 \to F'=4$ resonance peak of $^{85}$Rb, the laser remains stable for at least $30$ minutes. In this case, a rough estimation of the laser frequency fluctuations is obtained by considering that the error signal reveals a normal distribution which translates into a laser frequency standard deviation of the order of \unit{290}{\kilo\hertz}.

With the aim of establishing an absolute frequency scale for these measurements, which in turn allows for an all-optical determination of the hyperfine splitting and the hyperfine constants of the $6p_{3/2}$ level, a beam from the $\unit{420}{\nano\metre}$ laser is sent into a light modulation system based on an acousto-optical modulator (AOM) in a double pass configuration \cite{Donley2005}. The frequency modulated beam is sent through and retro-reflected into the spectroscopy system. This generates two new identical sets of peaks, both with a known shift with respect to the features produced by the original beam. This then provides the absolute frequency reference for the measured spectra.


\begin{figure}
\centering
\includegraphics[width=1\textwidth]{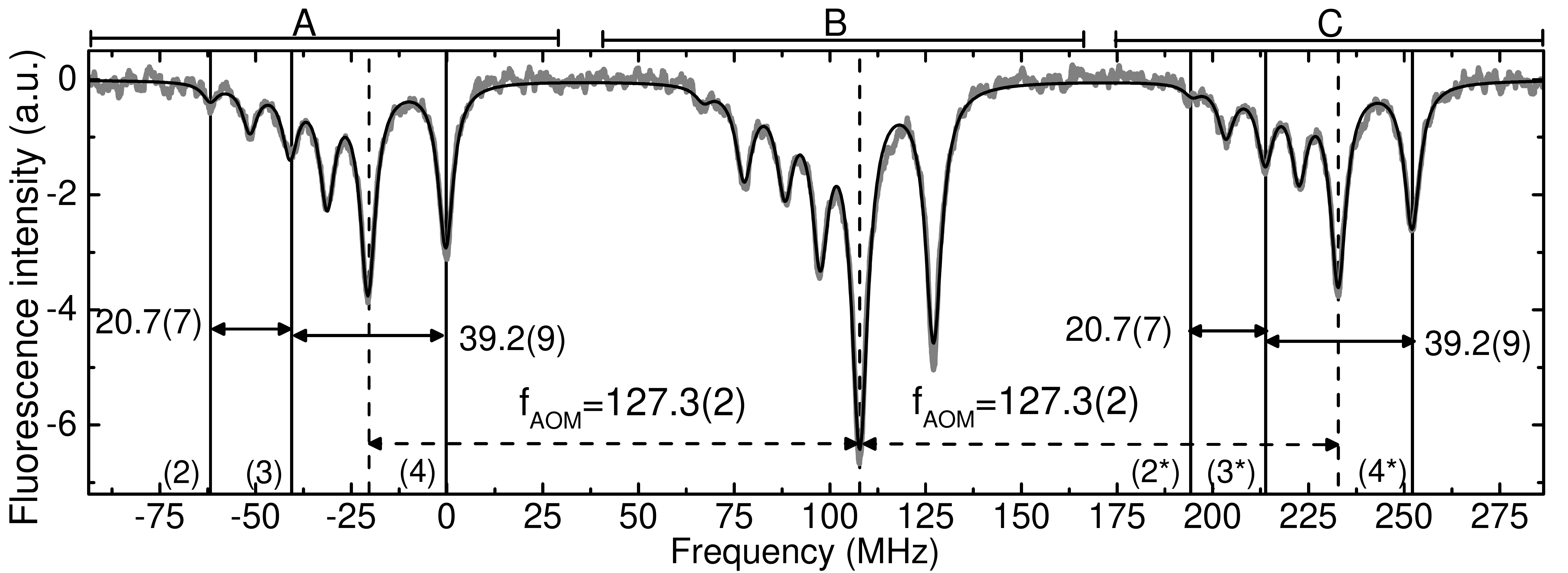}
\caption{Spectra for the transition\small{\textit{ $^{85}$Rb $5s_{1/2} F=3 \rightarrow 6p_{3/2} F'=4,3,2$}}. The spectra was recorded with the principal and modulated beam inside the Rb cell generating a set of three hyperfine resonances denoted by letters A, B and C.}
\label{picosFluorescenciaSat1}
\end{figure}
 
 


The spectrum shown in figure \ref{picosFluorescenciaSat1} is divided in three parts, denoted as A, B and C in the graph. 
Part A corresponds to the suppression of fluorescence due to  simultaneous interaction of the atomic vapor with two  co-linear and counter-propagating components of the original beam, both scanning with the same frequency and coming directly from the main laser without going through any additional modulation.   
On the right-hand side of the graph, part C of the spectra corresponds to fluorescence dips caused by the interaction of the thermal atoms with two co-linear and counter-propagating beam components of the modulated laser radiation.These beam components have a constant frequency modulation of $2\times f_{AOM}=2\times \unit{127.3(2)}{\mega\hertz}$ with respect to the original beams frequency. 
Finally, in the center of the graph, part B shows the fluorescence dips induced by two pairs of  co-linear and mutually counter-propagating components of the principal and the modulated beams simultaneously interacting with two separate and independent velocity groups of atoms, \emph{i.e.} two groups of atoms with velocities such that are resonant with components of the original and modulated beams.
Note that in this last case there are two velocity groups with opposite velocity directions that fulfill the requirement of being simultaneously resonant with light from both the original and modulated beams. These two groups independently contribute to the depth of the dips, which then almost doubles that of the suppressed fluorescence observed in parts A and C.  

Three main steps were taken to determine the hyperfine splittings and constants of the $6p_{3/2}$ level. Firstly, the Doppler background was subtracted from the recorded spectra. To achieve the background free spectra of figure \ref{picosFluorescenciaSat1}, it was necessary to use two gaussian profiles, each of which related to the Doppler broadened fluorescence signal independently caused by the principal and modulated beams. These two gaussian profiles have the same Doppler width but are shifted from each other by the AOM modulation frequency.

The second step is to perform Lorentzian fittings of each line. The fitting for the $18$ peaks was taken keeping the same line width for each group of peaks (group A, or B or C), but different between each group. The line width differences were mainly $2.8 \%$ and $7.5\% $ between AB and AC groups, respectively. After gathering the relative positions of their centers in the spectra, an average of the distances between the group peaks $\left(B_i-A_i\right)$ and $\left(C_i-B_i\right)$ was taken, with $i$ as the peak number from $1$ to $6$ of the hyperfine resonances and crossovers observed in each of the three parts of the spectra. All these differences correspond to the modulation frequency $f_{AOM}$. The average is taken to reduce discrepancies that can be attributed to the non-linearities of the PZT scan \cite{Jung2000}, instabilities of the laser, and uncertainties associated to the determination of the line centers, which in turn are the main limiting factors of the precision for the determination of the hyperfine structure of the $6p_{3/2}$ state presented in this work.
%
%
%

Next, with the calibration factor derived from the previous step and considering all the atomic resonances and crossover combinations of all three groups, together with ten spectra taken for each hyperfine ground state, a total of $90$ measurements for each hyperfine energy splittings of the $6p_{3/2}$ state were take into account to obtain the values shown in table \ref{tabla3}.
Finally, the hyperfine constants values can be found using the energy splittings and applying the hyperfine energy formula \cite{Arimondo}. An overdetermined equation system is obtained and resolved with a least squares regression. The resulting values are shown in table \ref{tabla4}.



  \begin{table}
  	\begin{center}
  		\begin{tabular}{@{} c c c c c @{}}
  			\toprule
  			 & {\bf Levels }  & {\bf This work }  & {\bf \citet{Bucka1961} }  \\
  			 &  ($F_2$,$F_1$)  &  $\Delta E$ (\mega\hertz) & $\Delta E$ (\mega\hertz) \\
  			\midrule
  			$^{85}$Rb \\
  			       & $(4,3)$ & $39.2(9)$ & $39.275(48)$  \\
  			       & $(3,2)$ & $20.7(7)$ & $20.812(61)$   \\
  			       & $(2,1)$ & $9.8(5)$ &  $9.824(44)$  \\
                   \cmidrule{1-4}
                   $^{87}$Rb \\
  			       & $(3,2)$ & $86.1(3.1)$ & $87.122(43)$  \\
  				   & $(2,1)$ & $50.8(1.5)$ & $51.418(31)$  \\
  				   & $(1,0)$ & $24.1(1.1)$ & $23.696(97)$  \\
  			  			\bottomrule	
  		\end{tabular}
  		\caption{Experimental and reported $6p_{3/2} $ hyperfine separations for the different energy levels of both Rb isotopes.}
        \label{tabla3}
  	\end{center}
  \end{table}


          \begin{table}
	\begin{center}
		\begin{tabular} {@{} r l l c l @{}}
			\toprule
			 &\multicolumn{2}{c}{\bf Hyperfine constants}  & \phantom{a} & \multirow{2}{*}{\bf Reference} \\
			\cmidrule{2-3}
			& $A$ (\mega\hertz) & $B$ (\mega\hertz) &&  \\
			\midrule
		$^{85}$Rb \\
			& $8.2(2)$  &  $8.2(6)$ && This work \\
  			& $8.179(12)$ & $8.190(49)$ && \citet{Arimondo} \\
  			& $8.220(3)$ &  $5.148(3)$ && \citet{ZhangConf2017}\\
\cmidrule{1-5}
			$^{87}$Rb \\
& $27.4(1.1)$ & $3.7(9)$ && This work \\
& $27.700(17)$ & $3.953(24)$ && \citet{Arimondo} \\
& $27.03$ & $3.772$ && Safronova \emph{et al.} \citep{Safronova2011} \\
  			\bottomrule	
		\end{tabular}
		\caption{Experimental hyperfine constants for both rubidium isotopes. The last column shows also the previous reported values.}
        \label{tabla4}
		\end{center}
	\end{table}


The experimental values of the hyperfine splittings for each of the four measured manifolds shown in table \ref{tabla3} are, within the precision of the measurement presented in this work and in agreement with the corresponding energy splittings reported by \citet{Bucka1961}. 
%
%
The hyperfine constants derived from the measured splittings presented in table \ref{tabla4} for both Rb isotopes are also in good agreement with the values previously reported in \cite{Bucka1961,Safronova2011} and recommended by \citet{Arimondo} as the most reliable hyperfine structure parameters for the $6p_{3/2}$ level. The latest and most precise measurement of the hyperfine constants of $6p_{3/2}$ level for $^{85}$Rb \cite{ZhangConf2017} is also included.
 Even though the precision reported by the authors of this recent publication is approximately one order of magnitude better than previous measurements, there is a strong ($37\%$) discrepancy with respect to the values recommended by \citet{Arimondo}, with which our measurement fully agrees. The authors of \cite{ZhangConf2017} acknowledge the difference but do not present an explanation for such disagreement. The Safronova reported values \cite{Safronova2011} are theoretical results calculated using the relativistic all-order method including partial triple excitations. Our results are in agreement with the reported values of this high-precision systematic study.  

In our results, the uncertainties remain above a few $\mega\hertz$. 
The main limiting factor in our measurements is due to non-linearities of the PZT scanning, mostly for broad scans $\sim\unit{300}{\mega\hertz}$. A good example of this is the measurement for $^{87}$Rb, for which the energy splittings are larger and so is the relative error. It is also important to consider that, as specified by the manufacturer, the linewidth of the laser is $<\unit{140}{\kilo\hertz}$. 
Nevertheless, the discrepancies with previous and more precise measurements reported in the literature are well within the uncertainties achieved by the spectroscopy system reported in this work.

 

\section{Conclusions} \label{Conclusions}

We present a simple experimental system for a sub-Doppler resolution in rubidium atoms at room temperature. This system allows to resolve the hyperfine structure of the excited $6p_{3/2}$ state of either of the two most abundant rubidium isotopes and has the advantage of resolving the atomic and crossovers transitions.

A numerical calculation for a rate equation model for the populations of  $^{87}$Rb and  $^{85}$Rb was explored. The calculation considers the entire hyperfine structure of all the involved states to describe the saturation effects on the cascade fluorescence. Also, we notice that there are effects of light polarization that tend to suppress the crossover transitions (2,1), (1,0) for the $^{85}$Rb ($F=2$) and $^{87}$Rb ($F=1$) respectively. We are currently analyzing these polarization effects and the absence of the aforementioned  crossover transitions. 

One clear advantage of our experimental scheme is the detection of fluorescence in a region of the spectrum that is far away from the excitation laser line. This simplifies the detection due to the fact that all scattered laser light is not sensed by the PMT. The geometrical simplicity of the detection system allowed the incidence of the principal and modulated beams inside the cell and thus an absolute frequency scale was achieved. 
This allowed the optical measurement of the hyperfine energy splittings with megahertz accuracy and in good agreement with previously reported values.\\

\begin{acknowledgments}
We thank J. Rangel for his help in the construction of the $1050 $ nm diode laser. This work was supported by DGAPA-UNAM, M\'exico, under projects PAPIIT Nos. IN116309, IN110812, and IA101012, by CONACyT, M\'exico, under Basic Research project No. 44986 and National Laboratory project LN260704. LM Hoyos-Campo thanks UNAM-DGAPA and Conacyt for the postdoctoral fellowship.
\end{acknowledgments}

\bibliography{Bibliografia_Fluores}




\end{document}